\documentclass{aastex61}

\usepackage{amsmath}
\usepackage{graphicx}

\begin{document} 

\begin{center}

{\LARGE\bf Autoregressive Times Series Methods for Time Domain Astronomy} \\~\\

Eric D. Feigelson$^{1,2}$, G. Jogesh Babu$^{2,3}$,  Gabriel A. Caceres$^1$ \\

$^1$ Department of Astronomy and Astrophysics, 525 Davey Laboratory, Pennsylvania State University, University Park PA 16802 \\
$^2$ Center for Astrostatistics, 325 Thomas Building, Pennsylvania State University, University Park PA 16802 \\
$^3$ Department of Statistics, 325 Thomas Building, Pennsylvania State University, University Park PA 16802 \\

\vspace{0.5in}

{\large Frontiers in Physics, vol. 6, id. 80 (2018) \\
d.o.i. 10.3389/fphy.2018.00080 \\
Special Issue: Imagining the Future of Astronomy and Space Science \\ ~ \vspace{0.3in} \\
} 

\end{center}
\begin{abstract}
Celestial objects exhibit a wide range of variability in brightness at different wavebands.  Surprisingly, the most common methods for characterizing time series in statistics -- parametric autoregressive modeling -- is rarely used to interpret astronomical light curves.  We review standard ARMA, ARIMA and ARFIMA (autoregressive moving average fractionally integrated) models that treat short-memory autocorrelation, long-memory $1/f^\alpha$ `red noise', and nonstationary trends.  Though designed for evenly spaced time series, moderately irregular cadences can be treated as evenly-spaced time series with missing data.  Fitting algorithms are efficient and software implementations are widely available. We apply ARIMA models to light curves of four variable stars, discussing their effectiveness for different temporal characteristics.  A variety of extensions to ARIMA are outlined, with emphasis on  recently developed continuous-time models like CARMA and CARFIMA designed for irregularly spaced time series.  Strengths and weakness of ARIMA-type modeling for astronomical data analysis and astrophysical insights are reviewed.  
\end{abstract}

\section{The variability of cosmic populations}

Except for five roving planets and an occasional comet or nova, the nighttime sky seems immutable to the human eye.  The pattern and brightness of stars appears unchanging as from our childhood to old age.  Myths from ancient Egyptian, Greek and Australian Aboriginal cultures suggest that a few stars (such as Algol, Mira and Aldeberan) were recognized as variables (Wilk 1996, Hamacher 2018). As telescopic studies proliferated from the 17th through 21st centuries, more variable stars were found with a wide range of characteristics.  Some are periodic due to pulsations, rotationally modulated spots, or eclipses of binary companions.  Others vary in irregular ways from magnetic flares, eruptions, pulsations, accretion of gas from companions, and most spectacularly, nova and supernova explosions.  Ten thousand stars in two dozen categories were catalogued by Kukarkin \& Parenago (1948); this catalog now has over 50,000 stars with $>100$ classes (Samus et al.\ 2017).  NASA's Kepler mission has recently shown that most ordinary stars are variable when observed with  $\sim$0.001\% accuracy and dense cadences (Gilliland et al.\ 2015).

The study of celestial objects with variable brightness has broadened hugely in recent decades, emerging as a recognized discipline called `time domain astronomy' (Griffin et al.\ 2012).  The brightest sources in the X-ray and gamma-ray sky are highly variable, typically from accretion of gas onto neutron stars and black holes.  Timescales range from milliseconds to decades with a bewildering range of periodic, quasi-periodic, stochastic and bursting characteristics. The Galactic black hole binary GRS 1915+105 alone has a dozen modes of variability (Belloni et al 2000).  The radio sky has extragalactic quasars and blazars as well as Galactic pulsars and several varieties of fast radio bursts and transients.  The non-photon gravitational wave observatories have recently emerged with rapid `chirps' from merging black hole and neutron star binaries (Abbott et al.\ 2016).  A huge industry searching for distant supernova explosions is propelled by their utility in tracing the accelerated expansion of the Universe (Riess et al.\ 1998).  

Given this tremendous growth in the volume and complexity of astronomical time series data, we can ask what methods are common for their characterization and analysis.  In the statistical literature, the dominant methods are time domain parametric modeling.  Popular introductory texts (Chatfield 2003, Brockwell \& Davis 2016) devote most of their chapters on autoregressive time domain models  with only brief treatment of Fourier frequency domain methods.  Prominent graduate level textbooks have a similar balance (Hamilton 1994, Shumway \& Stoffer 2017).  Yet only few research papers each year in the astronomical literature mention autoregressive time domain methods: $<10$ papers/year prior to 2010, recently growing to $\sim 25$ papers/yr.  In contrast, $\sim 1000$ papers annually use `Fourier' or `power spectrum' procedures.  

This radical dichotomy between the methodologies emphasized by statisticians and those used by astronomers  motivates our investigation here.  Perhaps the difference arises due to real differences in data characteristics and questions raised in a physical science like astronomy compared to questions raised in social sciences like econometrics that are treated in the statistical literature.  This seems possible: time series  texts oriented towards engineers (Percival \& Walden 1993, 2000) and meteorologists (Duchon \& Hale 2012) generally use spectral and wavelet analysis rather autoregressive modeling. Another difference is that social science time series mostly have evenly spaced observation times while astronomical time series are often irregularly spaced due to daily or annual celestial cycles and other causes.  The recent growth in autoregressive models for astronomy is mostly restricted to the narrow field of continuous-time models that treat irregular time series (Kelly et al.\ 2009).  

Our goal here is to investigate whether standard time domain methods commonly taught and used in the social sciences can in fact be effective in characterizing astronomical time series, and whether this analysis path can lead  to valuable scientific insights.  

We start with a brief presentation of the mathematics of common time domain models,  model fitting and validation,  and implementation of modeling in public domain software.  We proceed with application of standard methods to the brightness variations of four stars using both evenly- and irregularly-spaced light curves based on the work of Caceres et al. (2018a,b).  We continue with an outline of various extensions to the common autoregressive models, including new developments of continuous-time models appropriate for irregularly-spaced time series.  We end with discussion on what astronomical insights can emerge from autoregressive modeling, and comments on the utility of this approach to future time domain studies in astronomy.

\section{The ARMA, ARIMA and ARFIMA Models} \label{models.sec}

In this section, we give a simplified presentation of standard material on parametric autoregressive modeling.  The suite of ARMA-related procedures was brought into prominence by the 1971 textbook of Box and Jenkins (Box et al. 2016 is the latest edition) and is sometimes known as the Box-Jenkins method.  Additional coverage appears in time series analysis texts like Chatfield (2003) and (in greater mathematical detail) Hamilton (1994), and econometric textbooks like Enders (2015), Greene (2017), and Hyndman \& Athanasopoulos  (2017).  The ARFIMA (also called FARIMA)  model is less common than the ARIMA model ($\sim 0.05$M vs.\ $\sim 2$M Google hits).  Its mathematical foundations are treated in monographs such as Palma (2007).  Autoregressive modeling is reviewed in the astronomical literature by Scargle (1981), Koen \& Lombard (1993), and Caceres et al.\ (2018a).  

As elsewhere in statistical analysis, it is wise to examine the data nonparametrically prior to parametric modeling.  The time series should be examined for its basic behavior with kernel density estimation (or similar nonparametric regression) smoothing to see trends.  Box-plots, histograms and quantile-quantile plots show the distribution of intensity values.  Most importantly here, the nonparametric autocorrelation function (ACF) gives the correlation between a series and time-lagged values of itself over its entire length. For an evenly-spaced time series $x$ with time stamps $t$, mean value $\mu$ and standard deviation $\sigma$, 
\begin{equation}
  ACF(k) = \frac{E[(x_t - \mu)(x_{t-k} - \mu)]}{\sigma^2}.
\end{equation}
In realistic cases, the population mean and standard deviation are unknown and must be estimated from the data.  
As with Fourier and wavelet transforms, the full information in a time series is reproduced in the ACF if an unlimited number of coefficients are used.  Theorems give the statistical distribution of the ACF for Gaussian white noise, so confidence intervals (such as $P=99$\%) can be constructed and hypothesis tests applied (Durbin-Watson and Breusch-Godfrey tests) to readily evaluate whether autocorrelation is present.  

When the ACF shows significant signals, current values of the time series depend on past values.  The time series violates the assumption of `independence'  in i.i.d. and many standard statistical procedures can not be accurately applied.    This is the situation where autoregressive modeling can be effective.   Commonly, two forms of dependency on past values are treated as a linear regression.  First, an autoregressive (AR) process has coefficients that quantify the dependence of current values on recent past values:
\begin{equation}
  x_t = a_1 x_{t-1} + a_2 x_{t-2} + \ldots + a_p x_{t-p} + \epsilon_t 
\end{equation}
where $\epsilon_t$ is a normally (Gaussian) distributed random error with zero mean and constant variance, $p$ is the order of the process (i.e., how many lags to model), and the $a$'s are the corresponding coefficients for each lag up to order $p$.  Second, a moving average (MA) process has coefficients that quantify the dependence of current values on recent past random shocks  to the system:
\begin{equation}
   x_t = \epsilon_t + b_1 \epsilon_{t-1} + b_2 \epsilon_{t-2} + \ldots + b_q \epsilon_{t-q} 
\end{equation}
where $\epsilon_t$ is the error term for the $t$-th time point, $b_i$ is the coefficient for each lagged error term up to order $q$.   Here the past $\epsilon$ noise values can be viewed as random shocks to the system, and the $b$ coefficients quantify the response to the shocks, $\epsilon$.   Adding these two equations together gives a combined ARMA(p,q) process.  Coefficients are estimated by standard regression procedures such as maximum likelihood estimation. 

The ARMA model has several assumptions including stationarity that requires constant mean and constant variance.  A stationary time series exhibits the same behavior at the beginning, middle and end of the dataset.  Violations of constant variance (homoscedasticity) can be treated with nonlinear ARMA-type models with volatility like GARCH. Violations of stationarity can be tested and regimes of different behavior can be found using change point analysis.   

Nonstationarity and variable mean values can sometimes be removed by fitting a global regression model such as a polynomial, but often an adequate detrending regression model cannot be found.  A flexible nonparametric procedure called $differencing$ can remove nonstationarity in many such cases.  Here one applies the backshift operator $B$ that replaces the time series $x_t$ by another $x^\prime_t$ consisting of the point-to-point difference in values:
\begin{equation} \label{backshift.eqn}
  x^{\prime}_t ~=~ x_t-Bx_t ~=~ x_t ~-~ x_{t-1}.
\end{equation}
The differenced time series can then be modeled as a stationary ARMA process, and the original time series with trends can be recreated by reversing or $integrating$ the differenced time series.  This combination of nonparametric differencing and integration with a parametric ARMA process is called the ARIMA(p,d,q) model where $d$ represents the number of differencing operations applied and typically equals one.   

Finally, and less intuitively, a fractional integrated procedure can be described by  
\begin{equation}
  (1-B)^d x_t = \epsilon_t
\end{equation}
where $d$ can be a real (non-integer) order of differencing and $B$ is the backshift operator defined above.  When combined with ARMA modeling, this gives an ARFIMA(p,d,q) process.  The advantage here is that fractional values of $d$ quantifies a long-memory dependency of $x_t$ on past values in a single parameter.  Mathematically, this fractional integration process corresponds to the Fourier behaviors known as $1/f^\alpha$ `red noise' where $f$ is the frequency and $\alpha$ is the slope of a power law fit to the low-frequency Fourier power spectrum (Palma 2007).   It can also be viewed as a binomial series expansion,

\begin{equation}
  (1-B)^d = \sum_{k=1}^\infty \binom{d}{k} ~ (-B)^k.
\end{equation}
The parameter $d$ is typically constrained to be in the range $0.0-0.5$ to define a stationary model.  Three commonly used parameters are related to each other as follows: $d = \alpha/2$ and $d=H-1/2$ where $H$ is the econometricians' Hurst parameter. 

In simple language, ARFIMA(p,d,q) is a time series model where the autoregressive and moving average components treat many short-memory dependencies on recent past values, and the fractional integration component treats both many forms of trend and a $1/f^\alpha$-type long-memory dependency.   Best fit parameters for ARMA -type coefficients are typically calculated with maximum likelihood estimation for different values of $p$, $d$ and $q$, called the $order$ of the model. The optimal order is then selected to optimize the Akaike Information Criterion (AIC), a penalized maximum likelihood measure that balances improvements in likelihood with increases in model complexity.

ARMA, ARIMA and ARFIMA models are attractive for astronomical time series analysis for various reasons.  First, the models are very flexible, successfully modeling an astonishing variety of irregular or quasi-periodic, smooth or choppy, constant or variable mean light curves.  Second, the dimensionality of the model is relatively low with a moderate computational burden of the numerical optimization.  In contrast, local regression methods, such as Gaussian Processing regression, may have thousands of parameters with $O(N^3)$ computational burden although some computational efficiency has been achieved (Forman-Mackey et al. 2017). Third, error analysis on the parameters naturally emerges through the likelihood regression analysis.  Fourth, they are extensible to situations involving multivariate time series, combinations of stochastic and deterministic behaviors, change points, and (moderately) irregular observation spacing.   Autoregressive modeling is not well-adapted to situations with strictly periodic variations (where the signal is compactly concentrated in Fourier power coefficients) or with sudden eruptive events (where the nonstationary amplitude is not  greatly reduced by differencing). 

Altogether, ARIMA-type models have proved to be useful to a very wide range of physical and human-generated dynamic systems in engineering, econometrics, Earth sciences, and other fields.  But despite these advantages,  non-trivial ARIMA models have been used only rarely in time domain astronomy for compact radio sources (Lazio et al. 2001), Mira stars (Templeton \& Karovska 2009) variability, and quasars (Kelly et al. 2014).  ARFIMA models appear in only a single study on solar flare activity (Stanislavsky et al. 2009).

\section{Testing the Model} \label{test.sec}

The first action that every analyst should take after ARIMA-type modeling is model validation based, in part, on residual analysis.  This should follow any regression procedure for two reasons.   First, even though the `best fit' has been obtained in a maximum likelihood sense, the entire model family may not apply to the dataset under study; this problem is called 'model misspecification'.  Second, the model may be correctly specified but its underlying mathematical assumptions may be violated.  For example, many regression procedures assume the errors ($\epsilon$ in equation 1) are `normal i.i.d.'; that is, they are independently and identically distributed following a Gaussian distribution.  

As with any regression analysis, the unsequenced values of model residuals can be tested for normality using the nonparametric Anderson-Darling or Shapiro-Wilks tests.  The Jarque-Bera test for normality examined the skewness and kurtosis of the residual distribution.  The Breusch-Pagen and White tests are used to test whether the residuals have constant variance or whether heteroscedasticity is present;  this is a violation of the `identically' requirement in i.i.d.  

A suite of specialized `time series diagnostics' have been developed by econometricians that use the time sequence of residuals to test model validity in various ways.  These are hypothesis tests for specific characteristics where the statistic has a known distribution for the null hypothesis of Gaussian white noise.  Thus quantitative probabilities for deviations from Gaussian white noise can be obtained under many circumstances.   

The most well-known of these diagnostics is the Durbin-Watson test of serial autocorrelation; i.e. whether the AR(1) coefficient of the residuals is consistent with zero.  The Breusch-Godfrey test generalizes this to test significance of AR($p$) where $p$ is specified by the user.  The Box-Pierce and  Ljung-Box test are `portmanteau' tests that address all values of $p$ simultaneously..  Other nonparametric tests (such as the Wald-Wolfowitz runs test that groups the residuals into a binary `+' and `$-$' variable) may also be effective in testing the existence of temporal structure in model residuals.  

Several statistics tests for nonstationarity in the residuals. A simple nonparametric correlation coefficient like Kendall's $\tau$ can test for overall correlation. Global regressions can reveal large-scale trends and local regressions smooth the residuals and reveal small-scale structure.  But it is possible for a time series to be nonstationary but without trend.  One of these conditions is called `unit root' where the time series does not revert to the original mean after experiencing a shock.  The physicists' random walk is an example of a nonstationary diverging process.  The augmented Dickey-Fuller and Kwiatkowski-Phillips-Schmidt-Shin (KPSS) statistics test for trend stationarity against a unit root.  

\section{Software implementations in R} \label{software.sec}

Software for autoregressive modeling is widespread.  In Python, {\it statsmodel.tsa} and related libraries have $\sim 70$ functions implementing AR, ARMA, ARIMA and VAR (vector autoregressive) model fitting.  The {\it kalmanf} function estimates ARMA models with exact maximum likelihood estimation using the Kalman filter.  The {\it regime\_switching} function implements a nonlinear Markov switching dynamic regression. Time series diagnostics, trend regressors and other ancillary functions are included.  

Matlab has  several dozen infrastructure functions for manipulating time series objects.  But its power is in the Econometrics Toolbox with extensive methodology (and tutorials) including: time series decomposition,  filtering, ARIMA regression with non-i.i.d. errors, ARMAX estimation with exogenous covariates, GARCH models, multivariate VAR, AIC model selection, state space modeling with the Kalman filter,  time series diagnostics, simulation and forecasting.  This Matlab software costs several tens of dollars for a student, several hundred dollars for a university researchers, and several thousand dollars for a commercial license.  

The largest collection of autoregressive software resides in the public domain {\it R} statistical software environment (R Core Team 2018).  Base-{\it R} software includes infrastructure for `ts' (evenly spaced) and `zoo' (irregularly spaced) time series objects as well as basic ARIMA and ARIMAX functionality. But the bulk of advanced time series methods are implemented in   some of the $>$11,000 add-on CRAN packages.  Mature and extensive CRAN packages for ARIMA-type modeling  include {\it acp}, {\it arfima}, {\it BAYSTAR}, {\it BigVAR}, {\it bsts}, {\it carfima}, {\it carx}, {\it cents}, {\it cts}, {\it dse}, {\it fGarch}, {\it FGN}, {\it FitAR}, {\it FitARMA}, {\it forecast}, {\it glarma}, {\it imputeTS}, {\it ltsa}, {\it MARSS}, {\it mclcar}, {\it mlVAR}, {\it MSBVAR}, {\it MTS}, {\it partsm}, {\it perARMA}, {\it portes}, {\it prophet}, {\it robustarima}, {\it rugarch}, {\it sparsevar}, {\it stochvol}, {\it stsm}, {\it TSA}, {\it tsdecomp}, {\it tsDyn}, {\it tseries}, and {\it vars}.   Recently, the continuous-time CARFIMA model has been implemented in CRAN package {\it carfima}.   Together these packages provide more than a thousand functions relating to autoregressive modeling.  The package capabilities are briefly listed and classified in the CRAN Task View on Time Series Analysis.   An elementary introduction to {\it R} for astronomical time series appears in the text by Feigelson \& Babu (2012). 

 Important diagnostics for goodness-of-fit for time series and model fit residuals (such as the Durbin-Watson, Bruesch-Godfrey, Box-Pierce, Augmented Dickey-Fuller, and KPSS tests) are available in CRAN packages such as {\it tseries} (Trapletti \& Hornik 2018) and {\it lmtest} (Zeileis \& Hothron 2002).   These tests are described in econometrics textbooks (e.g. Enders 2015, Greene 2017) and in Wikipedia. 

The ARIMA and ARFIMA fits in Figures~1-4 here were obtained using the {\it auto.arima} and {\it arfima} functions in {\it forecast}, one of the most popular CRAN packages that is describe in the ebook by Hyndman \& Athanasopoulos (2017).   The data manipulation and plotting was performed using base-R.  Benchmarking shows that the {\it auto.arima} and {\it arfima} codes provide  AIC-optimized maximum likelihood fits in $O(N)$ wall-time: $\sim 0.4$ CPU-sec for $\leq$1,000 datapoints, in 4~CPU-sec for 10,000 datapoints, and so forth.  This is sufficiently efficient for analyzing large astronomical surveys. However, the new {\it carfima} code (below) is much slower. 

\section{Application to Stellar Photometry} \label{4stars.sec}

{\bf KIC 005880320 (Figure~1)\footnote
{Figure details.  The light curves in the upper-left panels of Figures~1 and 2 are obtained from NASA's MAST archive center at Web page http://archive.stsci.edu/kepler/data\_search.  The quantity plotted is PDC flux in parts-per-million about the mean.  Known instrumental effects have been removed and the the quarters have been `stitched' together to a common mean. The HATS-2 light curve in Figure 3 was obtained from https://hatsouth.org/planets/lightcurves.html.  We use photometry from aperture 1 after Transit Filter Algorithm is applied to reduce instrumental and atmospheric variations.  The light curve in the upper-left panel of Figure~4 was obtained from ASAS's Web page http://www.astrouw.edu.pl/asas.  Photometry from different years were 'stitched' together without any adjustments, 
}.
} ~ We start our application of autoregressive models with a seemingly simple case: a magnetically quiet star observed nearly continuously for four years by NASA's Kepler mission with an evenly spaced 29.4~min cadence.  This solar-type star, also called Kepler 758, has four confirmed `super-Earth' planets with periods ranging from 5 to 20 days and transit depth ranging from 100 to 300 ppm (parts per million).  The periodic dips from the transiting planets are readily seen in periodograms based on the autoregressive model residuals (Caceres et al.\ 2018b).

Here we examine the appearance of the light curve from its original state through differencing, ARIMA and ARFIMA modeling (left panels).  Quantitative measures include the inter-quartile range (IQR) and the autocorrelation function (ACF) at each stage of the analysis (right panels).  Although the photometric behavior of the original light curve looks to the eye like near-random noise, it is highly autocorrelated with ACF(lag=1)$\sim$0.4 with memory extending to a day or more (upper right).  The differencing operator does not reduce the noise level (IQR is unchanged) but reduces the long-memory autocorrelation leaving behind a strong short-memory component with ACF(1)$\simeq -0.6$.  The maximum likelihood ARIMA fit required five coefficients but left behind significant autocorrelation over several lags.  But the ARFIMA fit with only two coefficients, despite a $d$ value close to zero, reduces the residuals to near-white noise.  

{\bf KIC 004276716 (Figure 2)}~  This star is a bright K3-type star that is observed for 14 of the 17 quarters of the Kepler mission with an unconfirmed candidate planetary transit with orbital period around 20 days.  It is much more magnetically active than the star shown in Figure 1, exhibiting photometric variations from large spots that emerge and disappear after a few 30~day rotation periods.   The differencing operator removed nearly all of the long-memory autocorrelation and reduced the overall IQR by a factor of 6, and a few outliers are revealed.  An ARIMA with order-6 improved the fit somewhat but the ARFIMA model with order-8 reduces the noise level several times again and eliminates all remaining autocorrelation.  Here a significant ($d=1.14$) long-memory component is present.  Note that the residual noise levels seems variable (heteroscedastic), so it may be worthwhile trying GARCH modeling.  

Caceres et al.\ (2018b) has modeled the full sample of $\sim$ 200,000 Kepler stars with ARIMA and ARFIMA, and find the behavior shown in Figures~1 and 2 is typical for most, but not all, of the stellar variations.  Generally, the broader family of ARFIMA models do a better job in reducing autocorrelation, although most of the noise is removed by ARIMA models.  Very simple models such as AR(1) or ARMA(1,1) that do not include detrending by differencing are not adequate to model most stellar light curves.  Fortunately for their scientific goals, Caceres et al.\ (2018a, 2018b) find that the periodic planentary transit signals are mostly not incorporated into the maximum likelihood  autoregressive fits, as they occupy only a small fraction of the data points.  Their Transit Comb Filter periodogram, designed to treat a box-like transit after the differencing operator is applied, is able to recover most of the Kepler confirmed planets and uncover new planetary cadidates.  

{\bf HATS-2b (Figure 3)}~ Ground-based surveys suffer two major cadence problems compared to space-based surveys: a given star is not observable for $\sim$6 months every year due to solar motion, and a given star is not observable for daylight hours from a single site.  The Hungarian Automated Telescope South network seeks to mitigate the latter problem with three telescopes on different continents allowing nearly continuous coverage during a 6~month season (Bakos et al.\ 2013). In ground-based surveys, most of the autocorrelated noise arises from uncorrected atmospheric variations at the telescope rather than intrinsic variability of the star.  Nonetheless, planetary transits can be recovered from ground-based light curves in a small fraction of cases.  HATS-2 is a magnetically active K-type star with a hot Jupiter in a 32~hour orbit (Mohler-Fischer et al.\ 2013). 

The ground-based light curve here is quite different in behavior from the space-based Kepler light curve.  The observing cadence is often dense with an observation every $\sim 5$ minutes, but irregular gaps in observations over the 6-month period are common.  Strong dips in brightness occur due to passing clouds or other atmospheric phenomena.  The irregular spacing is binned onto a fixed time grid; we arbitrarily choose the same 29.4~min time interval characteristics of the Kepler data.  The result is that the original $\sim 20,000$ observations fill $\sim 10,000$ evenly spaced time bins of which 76\% have missing entries.  The more typical ground-based surveys with only one telescope will have $>90$\% missing data.  

Here we find that the autoregressive modeling suppresses the strong outliers attributable to clouds, but does not improve the overall noise amplitude: IQR=0.013 mag (magnitude) in the original light curve and 0.011 mag in the ARFIMA residuals.  But the autocorrelation in the original light curve is nearly entirely removed by the ARFIMA fit, although the ARIMA fit was less successful.  Experimenting with different bin widths we find that increased bin sizes reduces the IQR somewhat but increases the residual ACF.  We are finding that searching for planetary signals in the autoregressive residuals using our Transit Comb Filter periodogram is approximately as sensitive as the standard search for signals in the original lightcurve using the Box Least Squares periodogram (Stuhr et al., in preparation). 

{\bf RR Hyi (Figure 4)}~  The final light curve examined here has $\sim 1,000$ irregularly spaced observations spread over $\sim 10$ years.  The star is RR~Hyi (= DV~Oct), a poorly studied quasi-periodic long-period variable star at a location near the southern celestial pole that is visible for most of the year.  The photometric observations are from the All Sky Automated Survey (Pojmanski 1997).  Most observations of RR~Hyi are separated by hours to a few days.  We bin the observations onto a fixed grid with arbitrarily chosen width of 5 days.  The resulting light curve for autoregressive analysis has 660 observations of which 38\% are missing.  

The original light curve is obviously strongly autocorrelated on many scales.  Most of the autocorrelation is removed by the differencing operation and the IQR noise level is reduced by a factor of three.  However, neither ARIMA nor ARFIMA fits give any significant improvement in either noise or autocorrelation.  This is a case where parametric autoregressive modeling is ineffective; only the nonparametric differencing detrending step has any useful effect.  The failure is likely due to the small dataset with many gaps that misses much of the short-memory autocorrelation of the stellar variations\footnote{It is interesting to compare our ARIMA treatment here to a time-frequency and wavelet treatment of a similar quasi-periodic variable star, RV Tau, by \citet{Thibaut05}.}

\section{Extensions to ARIMA} \label{extensions.sec}  

In addition to ARFIMA for 1/f$^\alpha$ noise processes, there are many mathematical variants of ARMA and ARIMA modeling.  They are typically developed for economics time series analysis and are described in econometric textbooks like Enders (2015), Greene (2017), and Hyndman \& Athanasopoulos  (2017).

{\bf Seasonality}  Many economic variables have periodic variations with a 12-month period.  These might be due to agricultural planting cycles, holiday sales cycles, or annual corporate reports.  These periodic components are added to autoregressive components, just as a star's rotational period or a planetary orbit might produce periodicities superposed on a stochastic flaring pattern.  

{\bf Exogeneous covariates} Here a deterministic component is added to a stochastic ARMA-type model.  It is typically a linear model applied to new variables over the same time intervals as the original time series.  A common  approach is the ARMAX($p,q$) model
\begin{equation}
x_t =  \sum_{i=1}^p a_i x_{t-i} ~+~ \sum_{j=1}^q b_j \epsilon_{t-j}  ~+~ \sum_{k-1}^r c_k y_{tk}
\end{equation}
where the first two terms are the usual ARMA model for the temporal variable $x$, and the third term has linear dependence on $r$ new temporal variables $y$.   The exogeneous covariates can be any function linear in the parameters: periodic, or some deterministic trend (e.g. polynomial), or some tabulated variables without a simple mathematical form.  To improve their modeling of exoplanetary transits hidden in stochastic variability, Caceres et al.\ (2018a) create a function with periodic box-like dips with unknown depth.  The advantage accrued here is that the transit depth is treated as just an additional parameter of the model with best fit value and confidence interval established by maximum likelihood estimation.  The use of exogeneous covariates thus allows an astrophysically important parameter to be obtained along with ARMA parameters of less astrophysical interest.  

{\bf Volatility} In the standard ARMA model, the autoregressive components are combined with a standardized Gaussian white noise component with constant variance, $\epsilon = N(0, \sigma^2)$.  But the noise can often be heteroscedastic where its variance is temporally variable.  A common approach are the autoregressive conditional heteroscedastic or ARCH models initiated by Engle (1982) and Bollerslev (1986) where the squares of the residuals $\epsilon^2$ is itself an autoregressive or ARMA process,
\begin{align}
\epsilon_t &= \nu_t \sqrt{h_t}  \nonumber \\
h_t &= d_0 ~+~ \sum_{i=1}^p d_i \epsilon^2_{t-i} ~+~ \sum_{j=1}^q e_t h_{t-j}
\end{align}

The generalized ARCH or GARCH model is slightly more complex with both AR and MA components for the disturbances to the original ARMA process.  GARCH models proved to be wonderfully effective for economic time series such as inflation and stock market variations; the Google search engine gives over million hits for `GARCH'. Many extensions have been developed: GARCH-M or GARCH-in-mean process; IGARCH or integrated GARCH; TGARCH or threshold GARCH; EGARCH or exponential GARCH; NAGARCH or nonlinear asymmetrical GARCH; COGARCH or continuous time GARCH; and dozens more.  Astronomical volatility can be seen in solar magnetic activity and its consequent space weather measures, but it is not clear that GARCH-like models are effective for modeling these time series.  

{\bf Regime switching}~~ An important class of nonlinear time series represents situations where the system jumps between different regimes of stationary ARMA-type relationships, such as high and low states seen in some astrophysical accretion systems.  The switching mechanism is controlled by a hidden variable; if this variable is first-order autoregressive, then we have a Markov switching model.  These models can be combined with GARCH and deterministic trends.  

{\bf State space modeling}~~ This is a broader approach to parametric modeling of complicated time series where autoregressive and other stochastic behaviors can be combined with linear or nonlinear deterministic trends or periodic components (Durbin \& Koopman 2012). Functional behaviors can be linear or nonlinear, and can involve derivatives of state variables.  Noise can be Gaussian or non-Gaussian, white or autoregressive.  State space models are defined hierarchically with one level describing relationships between 'state variables' defining the temporal behavior of the system and other levels describing how the underlying system relates to observables.  Heterscedastic measurement errors common in astronomical data, for example, can be treated in a natural fashion.  The coefficients of the model are calculated by least squares or maximum likelihood estimation, often through a recursive algorithm such as the Kalman filter.  

\section{Continuous time modeling for irregular astronomical time series} 

A  mathematical limitation of classical autoregressive modeling is the restriction to evenly spaced datasets.  However, astronomers need methods that are robust to irregularly spaced observing cadences limited by daily and annual celestial cycles as well as telescope allocation constraints.  The underlying signal is typically generated by a continuous process of complex, often nonlinear, physical mechanisms that are not fully understood such as accretion disks and magnetohydrodynamical turbulence.  Analysis is commonly based on Fourier analysis, but its underlying theorems apply only for the limited situation of a sinusoidal signal superposed on Gaussian white noise.  This model is often inadequate to characterize the often multi-scale nonstationary structure in the time series.  Finally, a variety of analysis techniques are needed to capture different properties in order to classify heterogeneous ensembles of time series, such as the 40 billion unevenly spaced light curves expected from the forthcoming Large Synoptic Survey Telescope. 

In the stellar light curve applications above, we show that a sufficiently dense cadence of irregular observations can be successfully binned or interpolated onto a fixed time grid with `missing data', allowing classical ARIMA-type modeling to proceed.  However,  a more effective approach may be generalizations of stochastic difference equations to stochastic differential equations.  These have been studied mathematically starting in the 1940s to model physical phenomena such as Brownian motion and econometric variables (Parzen 1962, Bergstrom 1990).  Just as ARMA processes play a central role in the representation of time series with evenly spaced observations, CARMA (continuous-time autoregressive moving average) processes play an analogous role in the representation of time series with continuous time parameter allowing direct treatment of irregular time sampling.  Many variants are studied: Brownian motion and fractional Brownian motion with long-range dependence, Ornstein-Uhlenbeck process with exponential damping,  L\'evy process with heavy-tailed and asymmetric time series.  The interest in CARMA-type processes is motivated mainly by the successful application of stochastic differential equation models to problems in finance.

In astronomy, a continuous autoregressive CAR($p$) process was first applied by Koen (2005).  A simple CAR(1) model, variously known as the Ornstein-Uhlenbeck process and damped random walk model,  was brought into use for modeling quasar light curves in irregular astronomical time series by Kelly et al.\ (2009) and generalized to CARMA($p,q$) by Kelly et al.\ (2014). The latter paper gives the power spectral density, and autocovariance function at lag $\tau$ for the CARMA model. Kelly's  software implementation called {\it carma\_pack} on Github provides a Markov chain Monte Carlo sampler, an adaptive Metropolis algorithm combined with parallel tempering, for performing Bayesian inference on these continuous time autoregressive moving average models.  This model has been used in dozens of astronomical studies, particularly for study of variability of quasars for which unevenly spaced observations are commonly available (e.g., Graham et al. 2015, Guo et al. 2017).

CARMA models have favorable mathematical properties for astronomical use, although the statistical theory of continuous stochastic processes can be difficult (Garnier \& Wang 2008).  Mathematically, the power spectrum density of a CARMA process is a sum of Lorentzian functions with centroid, widths, and normalizations of the Lorentzian functions as free parameters (Brockwell 2001). This provides a significant amount of flexibility for modeling complicated power spectra, such as accretion disk systems with quasi-periodic oscillations and red noise, suggesting that CARMA modeling may be applicable to many classes of astronomical variables. Moreover, CARMA models have a state space representation enabling the use of the Kalman filter for calculating their likelihood function.   The computational complexity of calculating the likelihood function for CARMA models can therefore scale linearly with the number of data points in the light curve, an advantage for application to massive time-domain data sets.   

The CARMA models account for irregular sampling and measurement errors, and are thus valuable for quantifying variability, forecasting and interpolating light curves, and variability-based classification. But they suffer the same limitations as ARMA($p,q$) for regularly spaced data: it assumes (weak) stationarity of behavior across the light curve (for instance, no trend in brightness) and only treats short-memory autocorrelation out to $max(p,q)$ time intervals.  ARFIMA, on the other hand, treats nonstationarity and a powerlaw model of long-memory behavior in addition to these elements.  

The mathematical development of continuous-time generalizations of ARIMA and ARFIMA have only recently begun; the effort is reviewed by Brockwelll (2014).  Theorems underlying CARIMA are presented by McElroy (2013) and for CARFIMA (also called FICARMA) by Brockwell \& Marquardt (2005) and Brockwell (2014).  A maximum likelihood estimation procedure for CARFIMA is developed by Tsai \& Chan (2005) with R public software implementation released by Tak \& Tsai (2017).     We encourage time domain astronomers to try modeling with their {\it carfima} R/CRAN package when faced with irregular light curves of variability that includes trend and long-memory `red noise' characteristics.  It will be particularly valuable to compare the performance of standard ARIMA with `missing data' (above) and continuous-time CARIMA on real and simulated astronomical light curves.

\section{Insights from Autoregressive Modeling} \label{insights.sec}

Most often, the goal of the econometrician is to forecast future values of the time series.  But the astronomer is rarely concerned about forecasting brightness of their target star or quasar.  (An exception is solar activity that affects our terrestrial environment.) The astronomer has other goals:

{\bf Identifying outliers}~  As we saw with the four stars examined above, data points that are dramatically different from most autoregressive residuals are not necessarily distinctive in the original time series.  Common astronomical procedures such as `repeated 3-sigma clipping' to remove outliers will be less effective than expected if autocorrelation is present.  CRAN package {\it forecast} provides some tools for outlier detection in the presence of autocorrelation.  

{\bf Identifying change points}~ While GRS~1915+105 with dozens of distinct modes of variability is exceptional, it is not uncommon that an astrophysical system will exhibit high and low states or other systematic changes in behavior.  Time series methods are available for identifying change points include classic CUSUM and F tests and newer PELT and CROPS algorithms. These tests are implemented in {\it R} through CRAN packages like {\it strucchange}, {\it changepoint}, {\it BCP}, {\it ECP}, and {\it CPM}.  

{\bf Characterizing $1/f^\alpha$ red noise}~  The characterization of long-memory processes such as $1/f$-type noise may be relevant to the understanding of stochastic astrophysical processes like turbulence or accretion.  But statistically, this can be biased.  The astronomers' common procedure of linear least squares regression on the log-spectral power density, known in econometrics as the Geweke-Porter-Hudak (1983) estimator, can be biased.  It depends on unguided choices of frequency cutoff, detrending, spectral tapering and smoothing.   A dozen other approaches to estimating $\alpha$ (or equivalently, $d$ or $H$) are in use with no consensus on a single best approach (Rea et al. 2013).  We have been surprised that time series with increasing spectral power at low frequencies can often be adequately modeled with short-memory low-dimensional ARIMA models without long-range power law components;  that is, it can be difficult to distinguish stochastic red noise from other behaviors.  Altogether, we urge caution in quantitative assessment of red noise in astronomical time series. 

{\bf Classifying time series}~  Increasingly, large ensembles of astronomical time series are collected from wide-field surveys without prior knowledge of the cause of variability in each object.  Considerable efforts are made to classify (often irregularly spaced) light curves to identify transients and variable types in Big Data collections of light curves using training sets and machine learning techniques (e.g. Bloom et al.\ 2012, Cabrera-Vives 2017, Castro et al.\ 2018, Faraway et al.\ 2016, Morii et al.\ 2016, Wright et al.\ 2017, Zinn et al.\ 2017).  Autoregressive modeling results may be useful features to help discriminate variable classes.  These might include the amplitude of AR(1) or MA(1) coefficients, the model order $p$ and $q$, the presence of a significant $d$ value in ARFIMA, and time series diagnostics of autoregressive model residuals. 

{\bf Understanding astrophysical processes}~ It is not uncommon that astrophysical theory of (magneto)hydrodynamic situations such as accretion disks characterize the behavior in terms of Fourier coefficients.  However, it seems reasonable that characterization of astrophysical models in terms of autoregressive coefficients should also be possible.  The potential for new insights from this direction of research is unknown. 

\section{Concluding comments} \label{comments.sec}

Astronomers have tended to use a narrow suite of time series methods, particularly Fourier analysis.  This is appropriate when the physical phenomenon is strictly periodic with roughly sinusoidal shape such as radial velocity of circular Keplerian orbits or acoustic oscillations.  The power spectrum concentrates periodic signals but not stochastic behaviors such as autoregression.  There is real danger that inaccurate or even incorrect inferences will be made when autoregressive behaviors are present but are ignored by simplistic methods that assume the interesting signal is embedded only in Gaussian white noise. For example, detection of sudden flare events with a `3-sigma' criterion typically assumes Gaussian white noise whereas the effective sample size is reduced for autocorrelated signals.  The Bayesian Blocks procedure (Scargle 1998) for detecting intensity changes in count data similarly assumes Poisson white noise.  

Over several decades, researchers in many fields have found that low-dimensional autoregressive ARIMA-type models that quantify dependence on past behavior are amazingly effective in modeling many forms of variability, far more flexibly than low-dimensional (e.g. global polynomial) models that involve dependence on time alone. Extensive and user-friendly software is available for ARIMA-type modeling, particularly in the public domain {\it R} and {\it Matlab} statistical software environments.  

Parametric modeling with any mathematical formalism is only useful for time series if the variability behavior is encompassed by the chosen mathematical model.  In statistical parlance, the model must be correctly specified.  But this constraint can be checked.  Once a mathematical family of models is chosen and fitted to the data by maximum likelihood estimation, diagnostic tests can be applied to confirm that the model is indeed successful; for instance, testing that the model residuals are simple Gaussian white noise.   Caceres et al.\ (2018b) find that the aperiodic variability seen in most stars observed by NASA's Kepler mission can be effectively treated by  ARIMA-type models.  

A particular limitation of autoregressive modeling is the insistence on evenly spaced observation times.   However, in our examination of stellar light curves above, we converted time series with irregular observation times into a sequence of regular intervals by simple binning with `missing data' in empty time slots.  Density estimation (smoothing) or imputation procedures can also be applied;  ARIMA modeling itself can be used for imputation, filling in missing data in astronomical time series.  An alternative approach is to use continuous-time autoregressive models such as CARMA that treats irregular observation times directly with continuous time models.  

Our emphasis here is on situations where a significant component of the variation exhibit autoregressive behaviors so that future values depend on past levels and changes.  These components could arise as nuisance effects such as atmospheric fluctuations at the telescope, or they could be intrinsic to the astrophysical systems under study.  While our treatment has concentrated on stellar photometric light curves, the methodology can readily be applied to other situations arising in astronomical research:  optical or X-ray brightness variations from accreting supermassive black holes (quasars and Seyfert galaxies);  complex signals in gravitational wave detectors; rapid atmospheric variations in ground-based active optics instruments; space weather induced background in satellite observatories; and so forth.  
 
It is important to recognize that the scope of modern time series methodology is vast, propelled by complex problems in econometrics, signal processing, engineering, and Earth science systems.   We have covered only a small portion of the field here.  For example, change point analysis could help in study of astronomical transients, noise can be removed from astronomical signals with wavelet or sparsity methods, and irregularly sampled astronomical time series can be smoothed with local regression procedures.  We encourage astronomers to be more adventuresome in investigating the utility of sophisticated time series methods, including ARIMA-type autoregressive modeling, to address problems in the burgeoning field of time domain astronomy.

{\it Acknowledgements: } We thank Andrew Stuhr (Penn State) for substantial assistance in the data analysis, and Joel Hartman (Princeton) for collaboration on the HAT South analysis.  This work is supported by NSF grant AST-1614690 and NASA grant 80NSSC17K0122. \\

\noindent {\bf References}

\noindent
Abbott, B.\ P., Abbott, R., et al., 2016, Observation of gravitational waves from a binary black hole merger, {\it Phys. \\\hspace*{0.3in} Rev.\ Lett.}, 116, \#061102 \\
Bakos, G.\ \'A., Csubry, Z., et al., 2013, HATSouth: A Global Network of Fully Automated Identical Wide-Field \\\hspace*{0.3in} Telescopes, {\it Publ.\ Astro.\ Soc.\ Pacific}, 125, 154\\
 Belloni, T., Klein-Wolt, M., M\'endez, M., van der Klis, M.\ \& van Paradijs, J., 2000, A model-independent analysis of \\\hspace*{0.3in} the variability of GRS 1915+105, {\it Astron.\ Astrophys.}, 355, 271-290\\
 Bergstrom, A. R.\ 1990, {\it Continuous Time Econometric Modelling}, Oxford University Press\\
 Bloom, J.\ S., Richards, et al.\ 2012, Automating discovery and classification of transients and variable stars in the\\\hspace*{0.3in}  synoptic survey era, {\it Publ.\ Astro.\ Soc.\ Pacific}, 124, 1175\\
 Bollerslev, T., 1986, Generalized Autoregressive Conditional Heteroskedasticity, {\it J.\  Econometrics}, 31 (3): 307-327\\
 Box, G. E., Jenkins, G.\ M., Reinsel, G.\ C.\ \& Ljung, G.\ M.\ 2016, {\it Time Series Analysis: Forecasting and Control}, 5th \\ \hspace*{0.3in} ed., Wiley\\
 Brockwell, P. J., 2001, L\'evy-driven continuous time ARMA processes, {\it  Annals Institute Stat.\ Math.}, 53, 113-124\\
 Brockwell, P. J. \& Marquardt, T. 2005, L\'evy-driven and fractionally integrated ARMA processes with continuous time \\ \hspace*{0.3in} parameter, {\it Statistica Sinica}, 15, 477-494\\
 Brockwell, P.\ J.\ 2014, Recent results in the theory and applications of CARMA processes, {\it Annals Institute Stat. Math.},\\ \hspace*{0.3in}  66, 647-685\\
 Brockwell, P.\ J.\ \& Davis, R.\ A., 2016, {\it Introduction to Time Series and Forecasting}, 3rd ed., Springer\\
 Cabrera-Vives, G., Reyes, I., F\"orster, F., Est\'evez, P.\ \& Maureira, J.-C., 2017, Deep-HiTS: Rotation invariant \\ \hspace*{0.3in} convolutional neural network for transient detection, {\it Astrophys.\ J.}, 836, \#97\\
 Caceres, G. A., Feigelson, E. D., Babu, G. J., et al.\ 2018a, submitted for publication\\
 Caceres, G. A., Feigelson, E. D., Babu, G. J., et al.\ 2018b, submitted for publication \\
 Castro, N., Protopapas, P.\ \& Pichara, K.\ 2018, Uncertain classification of variable stars: Handling observational GAPS \\ \hspace*{0.3in} and noise, {\it Astron.\ J.}, 155, \#16\\
 Chatfield, C. 2003, {\it The Analysis of Time Series: An Introduction}, 6th ed., CRC/Chapman \& Hall\\
 Duchon, C.\ \& Hale, R., 2012, {\it Time Series Analysis in Meteorology and Climatology}, Wiley\\
 Durbin, J. \& Koopman, S.\ J., 2012, {\it Time Series Analysis by State Space Methods},  2nd. ed, Oxford\\
 Enders, W., 2015, {\it Applied Econometric Time Series}, 4th. ed., Wiley\\
 Engle, R.\ F.,1982, Autoregressive Conditional Heteroscedasticity with Estimates of the Variance of United Kingdom \\ \hspace*{0.3in} Inflation. {\it Econometrica}, 50 (4): 987-1007\\
 Faraway, J., Mahabal, A., Sun, J., Wang, X. \& Zhang, L., 2016, Modeling light curves for improved classification, {\it \\ \hspace*{0.3in} Statistical Analysis \& Data Mining}, 9, 1-11\\
 Feigelson, E. D.\ \& Babu, G.\ J., 2012, {\it Modern Statistical Methods for Astronomy with R Applications}, Cambridge Univ \\ \hspace*{0.3in} Press\\
 Forman-Mackey, D., Agol, E., Ambikasaran, S.\ \& Angus, R.\ 2017, Fast and scalable Gaussian Process modeling with \\ \hspace*{0.3in} applications to astronomical time series, {\it Astron. J.}, 154, \#220\\
 Garnier, H.\ \& Wang, L.\ (eds.) 2008, {\it Identification of Continuous-Time Models from Sampled Data}, Springer\\
 Geweke, J., Porter-Hudak, S., 1983, The estimation and application of long memory time series models, {\it J.\ Time Series \\ \hspace*{0.3in} Analysis}, 4, 221-238\\
 Gilliland, R.\ L., Chaplin, W.\ J., Jenkins, J.\ M., Ramsey, L.\ W.\ \& Smith, J.\ C., 2015, Kepler mission stellar and \\ \hspace*{0.3in} instrument noise properties revisited, {\it Astron. J.}, 150, \#133\\
 Graham, M. J., Djorgovski, S. G., Stern, D., Drake, A. J., Mahabal, A. A., Donalek, C., Glikman, E. Larson, S. \\ \hspace*{0.3in} \& Christensen, E., 2015, A systematic search for close supermassive black hole binaries in the Catalina \\ \hspace*{0.3in} Real-time Transient Survey, {\it Mon. Not. Royal Astro. Soc.}, 453, 1562-1576\\
 Greene, W.\ H., 2017, {\it Econometric Analysis}, 8th ed., Pearson\\
 Griffin, E., Hanisch, H.\ \& Seaman, R., 2012, {\it New Horizons in Time-Domain Astronomy}, Proc. IAU Symp \#285, \\ \hspace*{0.3in} Cambridge Univ.\ Press\\
 Guo, H., Wang, J., Cai, Z., Sun, M.\ 2017, How Far Is Quasar UV/Optical Variability from a Damped Random Walk \\ \hspace*{0.3in} at Low Frequency?, {\it Astrophys. J.}, 847,\#132\\
 Hamacher, D.\ W., 2018, Observations of red-giant variable stars by Aboriginal Australians, {\it Austr. J. Anthropology}, \\ \hspace*{0.3in} arxiv.org/1709.04634\\
 Hamilton, J.\ D., 1994, {\it Time Series Analysis}, Princeton\\
 Hyndman, R.\ J.\ \& Athanasopoulos, G.\ 2017, {\it Forecasting: Principles and Practice}, 2nd ed., ebook at otexts.org/fpp2.  \\
 Kelly, B.\ C., Bechtold, J.\ \& Siemiginowska, A., 2009, Are the variations of quasar optical flux driven by thermal \\ \hspace*{0.3in} fluctuations?, {\it Astrophys.\ J.}, 698, 895-910\\
 Kelly, B.\ C., Becker, A.\ C., Sobolewska ,M., Siemiginowska, A.\ \& Ulltey, P., 2014, Flexible and scalable methods for \\ \hspace*{0.3in} quantifying stochastic variability in the era of massive time-domain astronomical data sets, {\it Astrophys.\ J.}, 788, \\ \hspace*{0.3in} \#33\\
 Koen, C.\ \& Lombard, F., 1993, The analysis of indexed astronomical time series. Part One. Basic methods, {\it Mon.\ \\ \hspace*{0.3in} Not.\ Royal Astro.\ Soc.}, 263, 287\\
 Koen, C., 2005, The analysis of irregularly observed stochastic astronomical time-series - I. Basics of linear stochastic \\ \hspace*{0.3in} differential equations, {\it Mon.\ Not.\ Royal Astro.\ Soc.}, 361, 887-892\\
 Kukarkin ,B.\ V. \& Parenago, P.\ P., 1948, General Catalogue of Variable Stars, U.S.S.R. Academy of Sciences\\
 Lazio, T.\ J., Waltman, E.\ B., Ghigo, F.\ D., Fiedler, R.\ L., Forster, R.\ S.\ \& Johnston, K.\ J., 2001, A dual-frequency, \\ \hspace*{0.3in} multiyear monitoring program of compact radio sources, {\it Astrophs.\ J.\ Suppl.}, 136, 265-392\\
 McElroy, T. S.\ 2013, Forecasting continuous-time process with applications to signal extraction, {\it Annals Institute Stat.\ \\ \hspace*{0.3in} Math.}, 65, 439-456\\
 Mohler-Fischer, M., Mancini, L., Hartman, J., et al., 2013, HATS-2b: A transiting extrasolar planet orbiting a K-type \\ \hspace*{0.3in} star showing starspot activity, {\it Astro.\ Astrophys.}, 558, \#A55\\
 Morii, M., Ikeda, S., et al., 2016, Machine-learning selection of optical transients in the Subaru/Hyper Suprime-Cam \\ \hspace*{0.3in} survey, {\it Pub.\ Astro.\ Soc.\ Japan}, 68, \#104\\
 Palma, W.\ 2007, {\it Long-Memory Time Series: Theory and Methods}, Wiley\\
 Parzen, E.\ 1962, {\it Stochastic Processes}, Holden-Day\\
 Percival, D.\ B.\ \& Walden, A.\ T., 1993, {\it Spectral Analysis for Physical Applications}, Cambridge\\
 Percival, D.\ B.\ \& Walden, A.\ T., 2000, {\it Wavelet Methods for Time Series Analysis}, Cambridge\\
 Pojmanski, G., 1997, The All Sky Automatd Survey, {\it Acta Astronomica}, 47, 467\\
 R Core Team, 2018, R: A language and environment for statistical computing, R Foundation for Statistical Computing, \\ \hspace*{0.3in} Vienna AT. www.r-project.org\\
 Rea, W., Oxley, L., Reale, M.\ \& Brown, J., 2013, Not all estimators are born equal: The empirical properties of some \\ \hspace*{0.3in} estimators of long memory,  {\it Mathematics and Computers in Simulation}, 93  (10), 1016\\
 Riess, A.\ G., Filippenko, A.\ V., et al., 1998, Observational evidence from supernovae for an accelerating universe and \\ \hspace*{0.3in} a cosmological constant, {\it Astron. J.}, 116, 1009-1038\\
 Samus, N.\ N., Kazarovets, E.\ V., Durlevich, O.\ V., Kireeva, N.\ N., Pastukhova, E.\ N., 2017, General catalogue of  \\ \hspace*{0.3in}  variable stars: Verson GCVS 5.1, {\it Astronomy Reports}, 61 (1), 80-88\\
 Scargle, J.\ D., 1981, Studies in astronomical time series analysis. I - Modeling random processes in the time domain, {\it \\ \hspace*{0.3in} Astrophys.\ J.\ Suppl.}, 45, 1-71\\
 Scargle, J.\ D., 1998, Studies in Astronomical Time Series Analysis. V. Bayesian Blocks, a New Method to Analyze \\ \hspace*{0.3in} Structure in Photon Counting Data, {\it Astrophys. J.}, 504, 405-18\\
 Shumway, R.\ H.\ \& Stoffer, D.\ S., 2017, {\it Time Series Analysis and Its Applications with R Examples}, 4th ed., Springer\\
 Stanislavsky, A.\ A., Murnecki, K., Magdziarz, M., Weron, A.\ \& Weron, K., 2009, Modeling of solar flare activity from \\ \hspace*{0.3in} empirical time series of soft X-ray solar emission, {\it Astrophys.\ J.}, 693, 1877-1882 \\
 Tak, H.\ \& Tsai, H.\ 2017, carima: Continuous-time fractionally integrated ARMA process for irregularly spaced long-\\ \hspace*{0.3in} memory time series data, R package version 1.0.0, https://cran.r-project.org/web/packages/carfima/index.html\\
 Templeton, M.\ R. \& Karovska, M.\ 2009, Long-period variability of $o$ Ceti, {\it Astrophys.\ J.}, 691, 1470-78\\
 Trapletti, A. \& Hornik, K., 2018, tseries: Time Series Analysis and Computational Finance, R package v. 0.10-44, \\ \hspace*{0.3in} https://CRAN.R-project.org/package=tseries\\
 Tsai, H. \&  Chan, K.S., 2005, Maximum likelihood estimation of linear continuous time long memory processes with \\ \hspace*{0.3in} discrete time data, {\it J. Royal Stat.\ Soc., Ser.\ B}, 67(5), 703-16\\
 Thibaut, C.\ \& Roques, S.\ 2005, Time-scale and time-frequency analysis of irregularly sampled astronomical time series, \\ \hspace*{0.3in} {\it EURASIP Journal on Advances in Signal Processing}, 10, 1155\\
 Wilk, S.\ R., 1996, Mythological evidence for ancient observations of variable stars, {\it J. Amer. Assoc. Variable Star \\ \hspace*{0.3in} Observers}, 24 (2), 129-133\\
 Wright, D.\ E., Lintott, C.\ J., et al.\ 2017, A transient search using combined human and machine classifications, {\it Mon.\ \\ \hspace*{0.3in} Not.\ Royal Astro.\ Soc.}, 472, 1315-23\\
 Zeileis, A. \& Hothorn, T., 2002, Diagnostic checking in Regression Relationships. R News 2(3), 7-10. https://CRAN.R-\\ \hspace*{0.3in} project.org/doc/Rnews/\\ 	
 Zinn, J.\ C., Kochanek, C.\ S., et al.\ 2017, Variable classification in the LSST era: exploring a model for quasi-periodic \\ \hspace*{0.3in} light curves, {\it Mon.\ Not.\ Royal Astro.\ Soc.}, 468, 2189-2205

\clearpage\newpage

\includegraphics[width=0.95\textwidth]{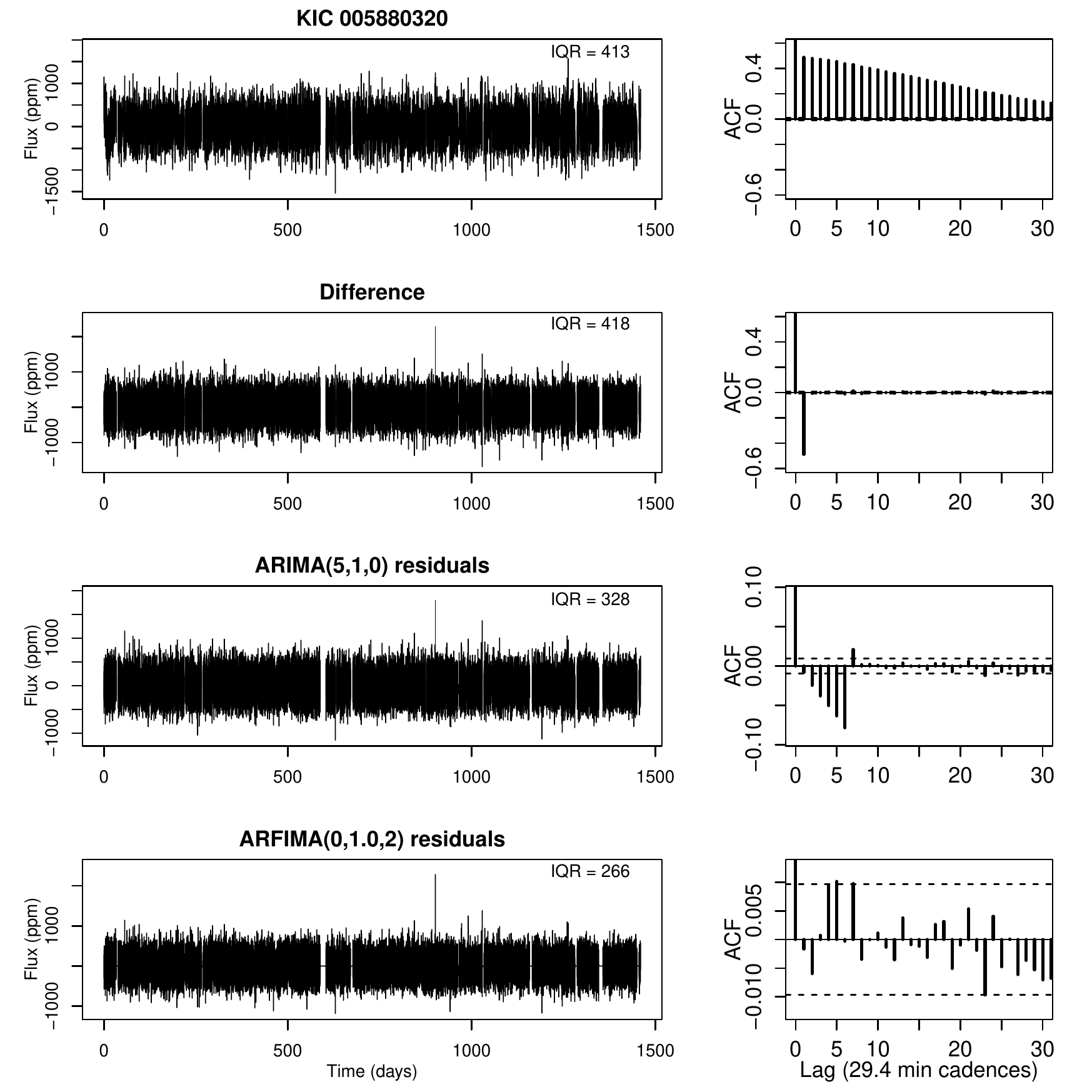} \\
Figure 1. ~~ Space-based Kepler mission photometric light curve of a quiet star with autoregressive modeling.   Left panels from top to bottom: Original light curve; differenced light curve; ARIMA model residuals; ARFIMA model residuals.   Right panels show the autocorrelation function at each stage.  \label{KIC_005880320.fig} 

\clearpage\newpage

\includegraphics[width=0.95\textwidth]{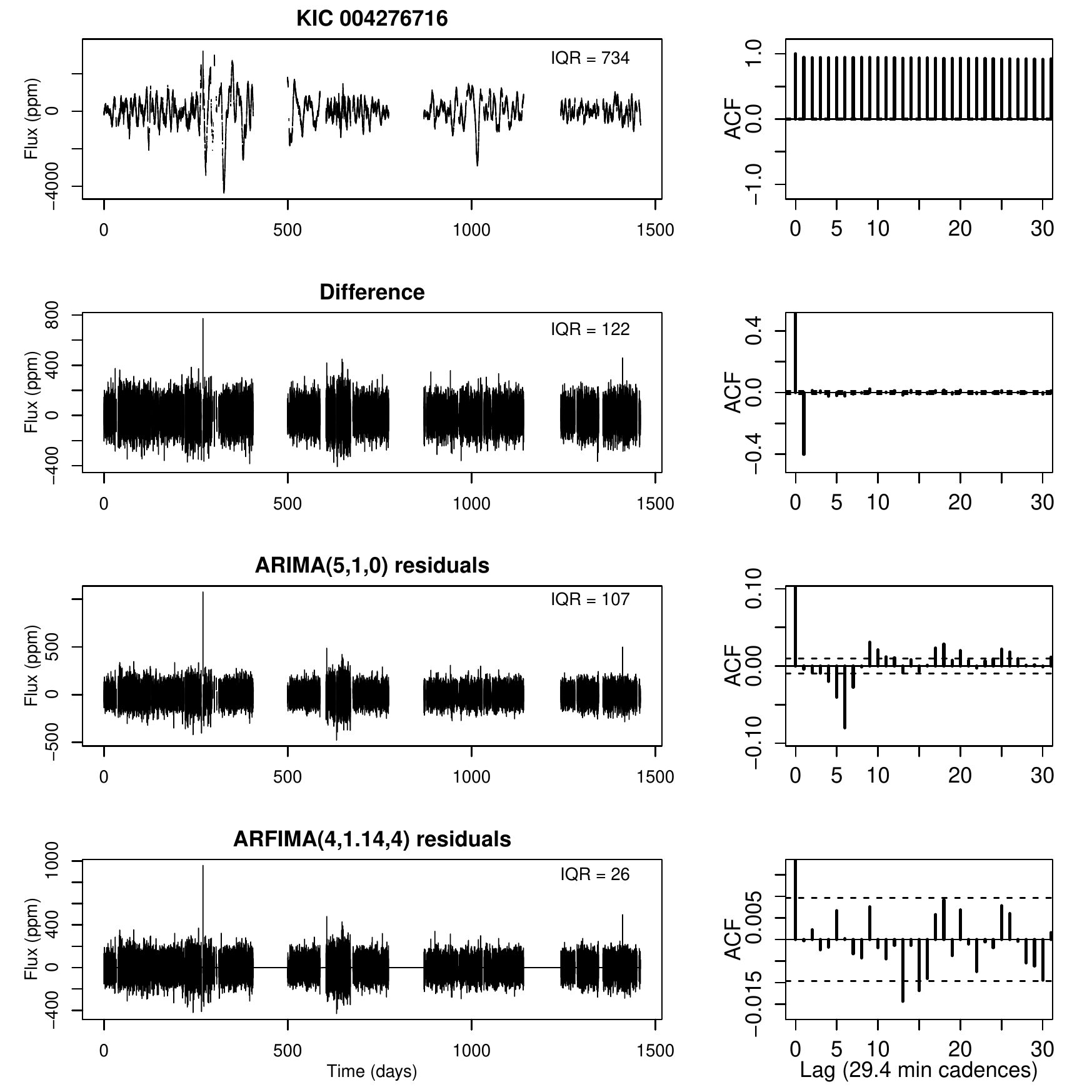}\\
Figure 2. ~~ Kepler mission light curve of a spotted  with autoregressive modeling.  \label{KIC_00426716.fig}

\clearpage\newpage

\includegraphics[width=0.95\textwidth]{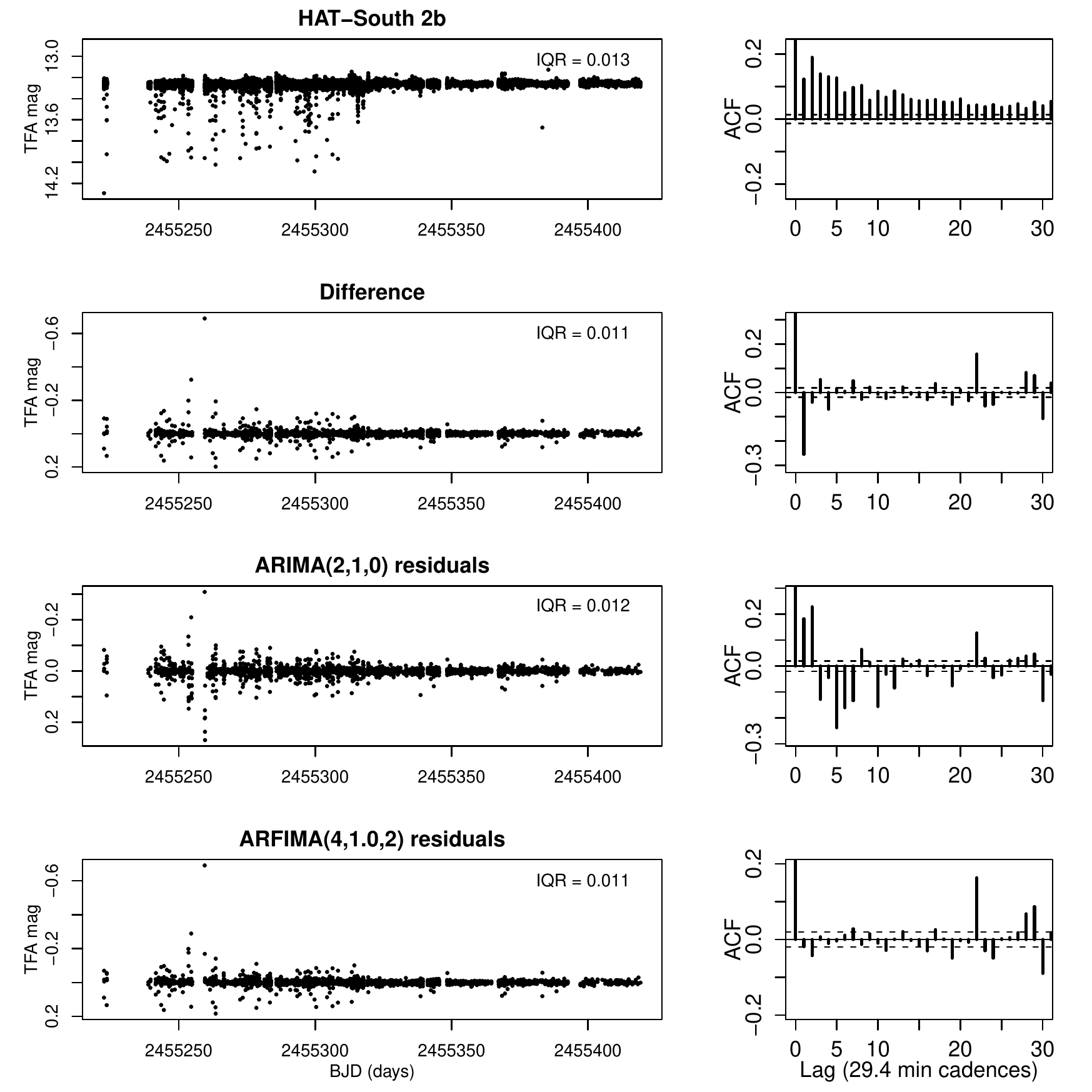}\\
Figure 3. ~~ Ground-based light curve for the planet-bearing star HATS-2b from the Hungarian Automated Telescope South network showing $\sim$6 months of brightness variations.  Except for the top panels, the irregular observation times have been binned onto a fixed sequence of 29.4 minute time bins.   \label{HATS_2b.fig}

\clearpage\newpage

\includegraphics[width=0.95\textwidth]{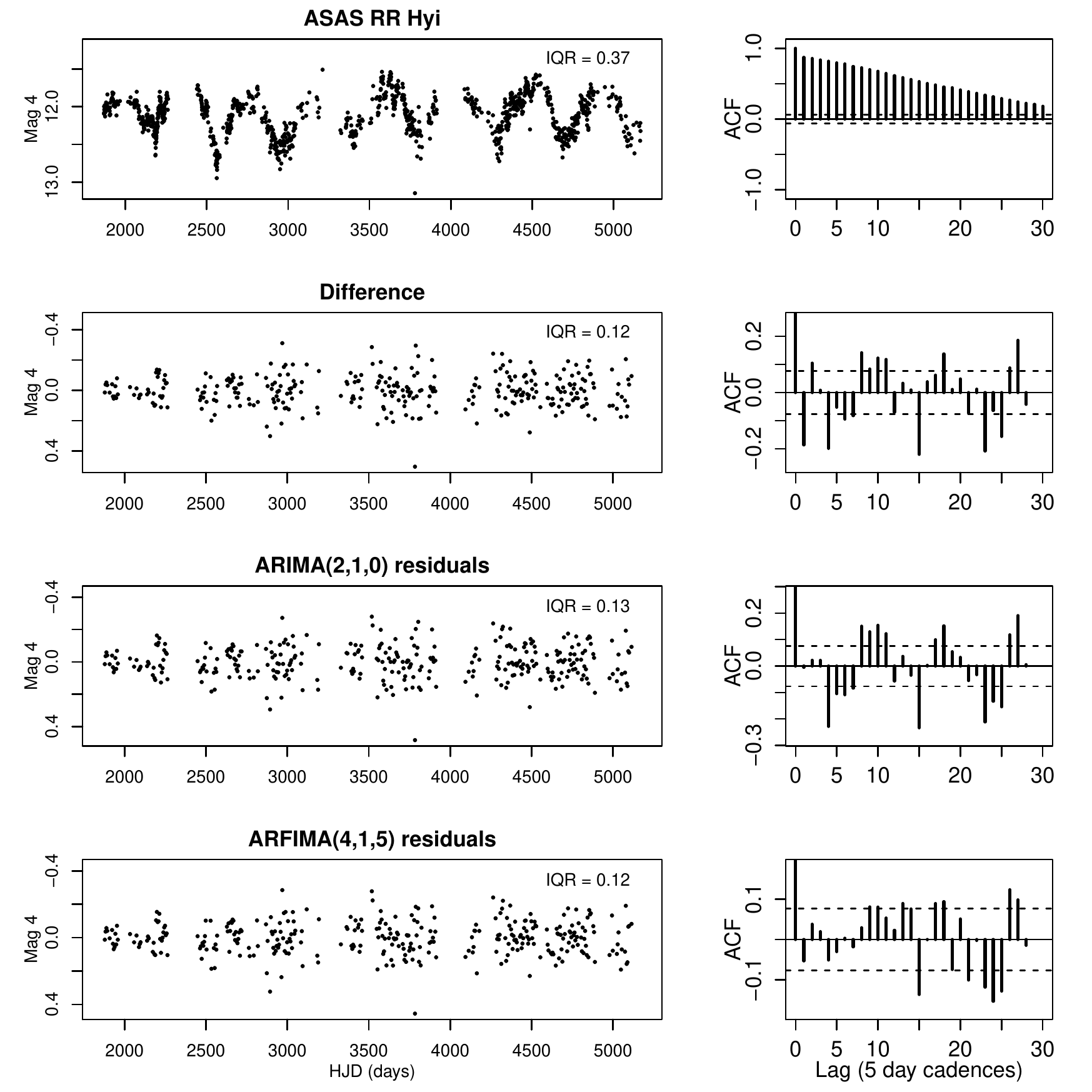}\\
Figure 4. ~~ Ground-based light curve for the variable star RR Hyi from the All-Sky Astronomical Survey (ASAS) showing $\sim$10 years brightness variations.  Except for the top panels, the irregular observation times have been binned onto a fixed sequence of 5 day time bins.  Left panels from top to bottom: Original light curve; differenced light curve; ARIMA model residuals; ARFIMA model residuals.   Right panels show the autocorrelation function at each stage.  \label{ASAS.fig}

\end{document}